\documentclass[namedreferences,hyperref,optionalrh,solaromanenum]{spr-sola}
\usepackage{bm}
\usepackage{CJKutf8}
\usepackage{amsmath}
\usepackage{comment}

\pdfoutput=1

\usepackage{graphicx}

\begin{document}

\begin{frontmatter}

\title{Investigating Pre-flare Signatures in Spectroscopic Observations of an X9-class Solar Flare}

%
\author[addressref={aff1,aff2,aff3}, email={ls593@njit.edu}]
       {\inits{L.}\fnm{Louis}~\snm{Seyfritz}\orcid{0009-0006-9335-6947}}

\author[addressref={aff3,aff4,aff5}, email={maria.kazachenko@colorado.edu}]
       {\inits{M.D.}\fnm{Maria}~\snm{Kazachenko}\orcid{0000-0001-8975-7605}}

\author[addressref={aff5}, email={ryan.french@lasp.colorado.edu}]
       {\inits{R.J.}\fnm{Ryan}~\snm{French}\orcid{0000-0001-9726-0738}}

%
\address[id=aff1]{Center for Solar-Terrestrial Research, New Jersey Institute of Technology, 323 M L King Jr Boulevard, Newark, NJ 07102-1982, USA}

\address[id=aff2]{Observatoire de Paris, Université PSL, CNRS, Sorbonne Université, F-92195 Meudon, FR}

\address[id=aff3]{National Solar Observatory, 3665 Discovery Drive, Boulder, CO 80303, USA}

\address[id=aff4]{Department of Astrophysical and Planetary Sciences,
University of Colorado Boulder, 2000 Colorado Avenue, Boulder, CO 80305, USA}

\address[id=aff5]{Laboratory for Atmospheric and Space Physics,
University of Colorado Boulder, 3665 Discovery Drive, Boulder, CO 80303, USA}

%
\runningauthor{Seyfritz et al.}

\begin{abstract}
    On October 3rd, 2024, the Sun emitted an X9.0-class flare from active region NOAA 13842. The event was recorded by multiple space-based instruments, beginning hours before the eruption, granting a unique opportunity to provide insight into the flare’s pre-flare phase.
    In this study, we employ analysis of Interface Region Imaging Spectrograph (IRIS) spectroscopic data to investigate pre-flaring phenomena associated with this flare.
    We present time-series and wavelet analysis of non-thermal velocity, Doppler velocity, and line intensity quantities of the IRIS Si IV 1403 \AA~line.
    We find two ranges of periodic oscillations during the pre-flare phase: $\sim$7-10 min and $\sim$18-21 min oscillations, with local enhancements occurring near the polarity inversion line.
    We also find a steady rise in Si IV line parameters beginning 3 hours before the flare in the same region, transitioning into strong non-thermal velocities and blueshifts $\sim15$ minutes before onset.
    These findings are consistent with a slow destabilization of the coronal magnetic field, possibly driven by the gradual activation of a flux rope, followed by a rapid shift to intense reconnection activity leading to flare onset.
\end{abstract}
\end{frontmatter}

\section{Introduction}\label{intro}

\begin{figure}
\centerline{\includegraphics[width=\textwidth]{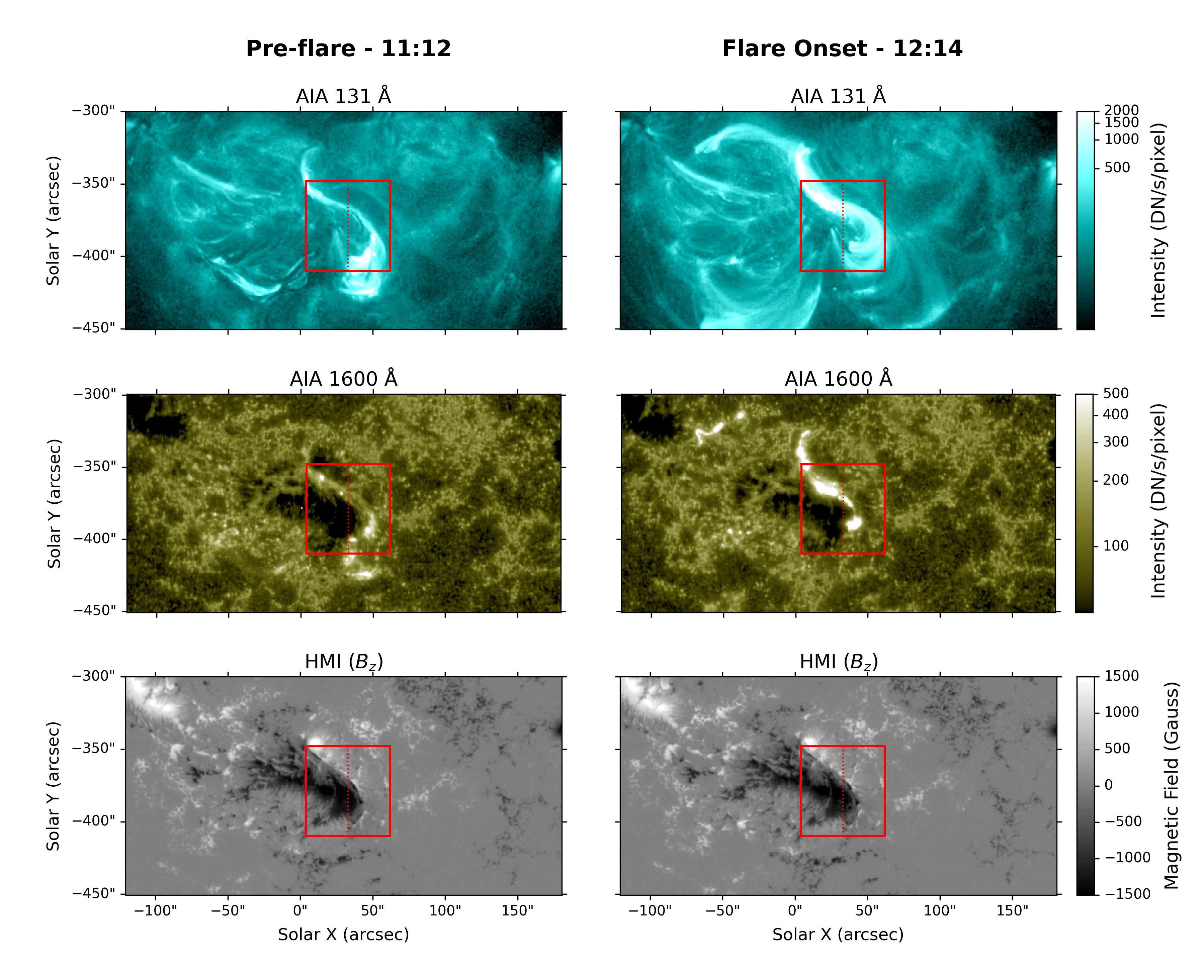}}
\small
\caption{Maps of 131 \AA\ and 1600 \AA\ AIA channels with HMI ($B_z$) measurements of active region 13842 on October 3, 2024. The left column shows snapshots $1$ hour before the flare at 11:12 UT, and the right column shows snapshots at 12:14 UT, one minute before the start of the X9.0 flare. The red frame corresponds to the SJI FOV and the dotted line to the fixed location of the spectrograph's slit.
See \S~\ref{observations} for more details.}
\label{cmaps}
\end{figure}

Solar flares evolve through a series of well-defined stages, particularly apparent in large events: a pre-flare phase, an impulsive phase, and a gradual phase \citep{Benz2008, Holman2011}.
The impulsive phase marks the onset of the flare, and is driven by the rapid release of stored magnetic energy into particle acceleration and emission over the whole electromagnetic spectrum \citep[e.g.][]{Fletcher2011}. This is followed by the gradual phase, during which the flare plasma cools, emitting predominantly soft X-rays \citep{Fletcher2008, Guo2011}.
Before these main phases, the pre-flare period plays a crucial role in the flare’s development. During this stage, energy slowly accumulates, and the magnetic configuration evolves towards a critical state. Although less energetic, the pre-flare period exhibits a variety of important physical and observational signatures \citep{Moore1992, Zhu2024}.

Pre-flare activity often manifests as Doppler shifts \citep[blueshifts,][]{Harra2013, Woods_preflaring}, localized brightenings, and structural changes in coronal loops \citep{Wang2017, Joshi2011}.
Localized heating during the pre-flare phase is frequently detected in EUV channels sensitive to hot coronal plasma \citep{Chifor2007}, such as those observed with the Atmospheric Imaging Assembly (AIA, \citealt{lemen2012aia}).
These observations indicate that energy is already being deposited into coronal loops or footpoints well before the flare’s impulsive phase begins \citep{Fletcher2011}.
The transition region is notably active during the pre-flare phase, responding dynamically to evolving energy deposition in the overlying corona. As coronal flare heating increases locally, thermal conduction and non-thermal particles can drive upflows and enhanced emission from the transition region. 
Observations from the Interface Region Imaging Spectrograph (IRIS, \citealt{depontieu2014iris}) have revealed transient brightenings and line broadening in transition region lines such as Si IV or Mg II, indicating both heating and plasma motion \citep{Woods_preflaring}. These features are also observed in the Fe XII coronal line in \cite{Woods_preflaring} and \cite{Harra2013} using the Hinode Extreme Ultraviolet Imaging Spectrometer (EIS, \citealt{Hinode_EIS}).
High-resolution imaging and spectroscopy have uncovered signatures of small-scale magnetic reconnection during the pre-flare stage \citep{Li2017}. These early reconnection events often occur at quasi-separatrix layers or within current sheets. They are typically observed as transient brightenings, jet-like ejections, or plasma flows, evidence of localized heating and acceleration. Such activity can destabilize the surrounding magnetic structure, and potentially trigger the large-scale reconnection that leads to the flare \citep{Li2016}.

Recently, \citet{To2025} presented an analysis of the pre-flare activity of 1,449 flares. Using a Hinode/EIS survey of flares from 2011–2024, \citet{To2025} found that non-thermal velocities at loop footpoints exhibit a systematic increase prior to the GOES (Geostationary Operational Environmental Satellite, \citealt{garcia1994goes}) recorded soft X-ray flare onset. For C–M class events, the typical lead time of these increases is a few to a few tens of minutes (4-25 minutes).

Another recorded feature of the pre-flare phase is the occurrence of oscillations (e.g., in intensity or velocity). 
Observational studies report both longer (periods of order 8–30 min) and shorter (minute-scale) period oscillations occurring minutes to hours before flare onset, identified in the time series of GOES SXR, H$\alpha$, EUV and microwave data \citep{Sych2009, Tan2016, Li2020}. 
These pre-flare oscillations have been interpreted as signatures of MHD waves, current‐loop oscillations, or transverse motions that can modulate electric currents and magnetic stress in active regions \citep{Nistico2013}.
Detecting and characterizing these oscillations for different flare phases is crucial for understanding their role in building magnetic stress and triggering the reconnection processes leading to the flares.

In this study, we present an analysis of the five hours of observations before an X9.0 flare on October 3rd, 2024, to provide insights into pre-flare processes before large solar flares. 
The paper is structured as follows.
We first describe the observational data and the methods used to analyze the optical and spectral measurements in Section \ref{obs}. We then present our results from the spectral analysis of the Si IV line of IRIS in Section \ref{res}, discussing the implications of our temporal, spatial and wavelet findings in Section \ref{diss}, and summarizing our study in Section \ref{conc}.

\section{Observations and Methods}\label{obs}

\subsection{Observations}\label{observations}

\begin{figure}
\centerline{\includegraphics[width=\textwidth]{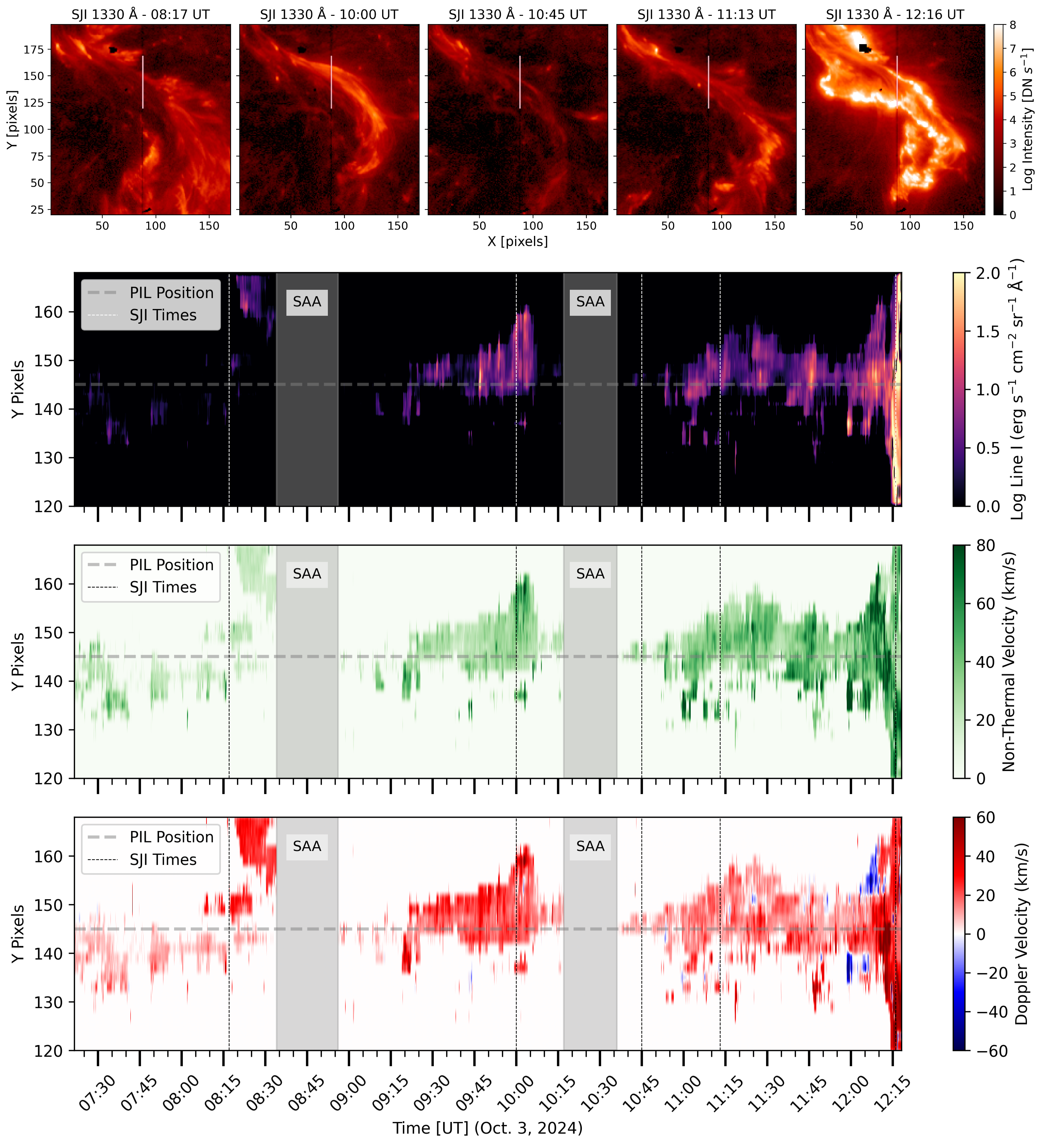}}
\small
\caption{\textit{Top row panels:} Overview of the pre-flare activity from IRIS/SJI (1330 \AA{}, Oct. 3rd 2024), covering up to five hours of pre-flare evolution before the X9.0 flare at 12:18 UT peak time. The first four images correspond to pre-flare phase, and the last one corresponds to the flare's onset. 
The vertical center line on the panels represents the slit, and in pink are the pixels chosen for spectral analysis in the maps below.
\textit{Rows 2-4:} Time-distance maps of logarithmic line intensity, non-thermal velocity, and Doppler velocity derived from the 1402.77 \AA\ Si~IV line along the pink portion of the IRIS slit (corresponding to the PIL). The vertical dotted lines on the lower rows mark times of the top row. See \S~\ref{obs} and \S~\ref{temp_res} for more details.}
\label{sji}
\end{figure}

The X9.0 class flare occurred on October 3rd, 2024, in NOAA active region (AR) 13842, starting at 12:15 UT and reaching its peak intensity at 12:18 UT. The flare was observed by SDO, IRIS, and GOES/XRS (see Figure \ref{cmaps} for context images). 
AR 13842 had already been very active the days preceding this flare, producing an X7.1 class flare on October 1, 2024, and an M-class flare at 8:15 UT of October 3rd, only 4 hours before the X9.0 flare.

We used observations from SDO and IRIS with varying cadences.
AIA captured full-disk images at a cadence of 12 seconds for the extreme ultraviolet channels (e.g. AIA 131 \AA), and 24 seconds for the ultraviolet channels e.g. 1600 \AA), with spatial sampling of ~0.6$^{\prime\prime}$ per pixel. The Helioseismic and Magnetic Imager (HMI, \citealt{schou2012hmi}) captured the vertical magnetic field maps at a 45 seconds cadence with a spatial sampling of ~0.5$^{\prime\prime}$ per pixel.
The IRIS observations of October 3rd were taken with a high-cadence sit-and-stare flare raster [OBS 4204700237], with the slit centered on the polarity inversion line
and a slit length of $62^{\prime\prime}$. The cadence was  $0.8$ seconds while the co-observed Slit Jaw Imager (SJI) 1330 \AA\ channel observed at an $8$-seconds cadence. The IRIS data analyzed spans from 07:21:27 UT to 12:18:00 UT covering five hours before the flare. Analyses of the impulsive flare phase of this IRIS dataset can be found in \citet{Kowalski2025,French2025b}.

We used full-disk AIA images in AIA 131 \AA~ and 1600 \AA~ channels to provide a larger field-of-view context to the IRIS observations.
Similarly, we used a photospheric line-of-sight magnetogram from HMI to identify the location of the polarity inversion line (PIL) of AR 13842. We used GOES X-ray flux measurements in the 0.5–4 \AA~ to derive global, disk-integrated view of soft X-ray emission and track early thermal signatures of pre-flare development (see Figure \ref{1D}).
In Figure \ref{cmaps}, we show the AIA 131 \AA{} and 1600 \AA{} observations along with the HMI line-of-sight magnetogram centered on the active region of interest, overlaid with the IRIS FOV (red square) and raster line location (dotted line).

In this study, we used observations of the optically-thin IRIS Si IV 1402.77\,\AA\ line, measuring plasma in the upper chromosphere and transition region ($T \sim 8 \times 10^4$\,K).
In the top row of Figure \ref{sji} we show the IRIS SJI 1330 \AA{} snapshots at four times during the pre-flare period, with the final frame from a minute after the flare start time.

\subsection{Methodology}\label{methods}

We track the evolution of IRIS Si IV spectra intensity, Doppler velocity, and non-thermal velocity in AR 13842, throughout the pre-flare period. 
Figure \ref{1D} (bottom panel) shows lightcurves from GOES 0.5-4 \AA, IRIS SJI 1330\AA{}, AIA 131 \AA\ and AIA 1600 \AA, starting $\approx5$ hours before the flare start time. For the AIA channels, the lightcurves are calculated over the FOVs presented in Figure \ref{cmaps}.
We analyzed the intensity variations from the lightcurves of these instruments to track the overall evolution of AR 13842 during the pre-flare phase. We used AIA intensity maps to track the evolution of AR 13842. 
We co-aligned AIA maps with HMI magnetograms to identify the PIL's location and corresponding IRIS slit pixels.

To analyze the Si IV 1403 \AA~ spectral profiles from IRIS, we used a modified version of the \texttt{fit\_siiv} function from the \texttt{iris\_lmsalpy} package available on GitHub. This routine fits the observed spectra at each spatial pixel using a Gaussian with linear background terms. From these fits, we extracted key physical parameters: peak intensity, Doppler velocity (relative to the rest wavelength of 1402.77 \AA~), non-thermal velocity, and line-integrated intensity. 
The non-thermal velocity ($v_{\rm nt}$) represents the portion of the spectral line’s broadening that remains after removing instrumental and thermal broadening, and reflects unresolved motions likely due to turbulence or waves.
The fitting was performed over a defined wavelength range, using constrained initial guesses to ensure stability, and filtering out unreliable results such as singular-valued or poorly fitted peaks.

We applied a factor of 2 spatial binning (decreasing from 208 to 104 pixels) and a factor of 10 temporal binning to improve signal-to-noise ratio of Si IV data. We then used it to derive line intensity, Doppler velocity and non-thermal velocity and thus focus on their  broader temporal trends, as shown in Figure \ref{sji}. We focused our analysis on a subset of $24$ spatial rebinned pixels corresponding to the S-shaped structure (or sigmoid) along the PIL, which remained visible throughout the pre-flare phase. 
We used this region to track the evolution of intensity, Doppler shifts, and broadening across the 4 hours 57 minutes of IRIS data before the flare.

To identify transient oscillatory features in the pre-flare lightcurves, we applied a continuous wavelet transform (CWT) with the Morlet wavelet method \citep{Wavelets_theo}.
We used a Savitzky-Golay filter \citep{savitzky1964smoothing} to detrend the signal over a 45-minute window prior to wavelet analysis, isolating residual fluctuations. The wavelet power spectrum was computed as $|\mathcal{W}(\tau, t)|^2$ (the wavelet transform in scale and time) and averaged in time to produce a global wavelet spectrum (GWS). Significance levels were determined by comparing power to a power-law noise background \citep{Wavelets_theo}. We only interpreted periods and times within the cone of influence and above the confidence threshold. This method allowed detection and validation of localized periodic behaviors.

\section{Spectral Analysis of Si IV: Doppler Velocity, Non-thermal Velocity, and Line Intensity}\label{res}

\subsection{Evolution During Early Pre-flare Phase (5 Hours Pre-Flare)}\label{temp_res}

\begin{figure}
\centerline{\includegraphics[width=\textwidth]{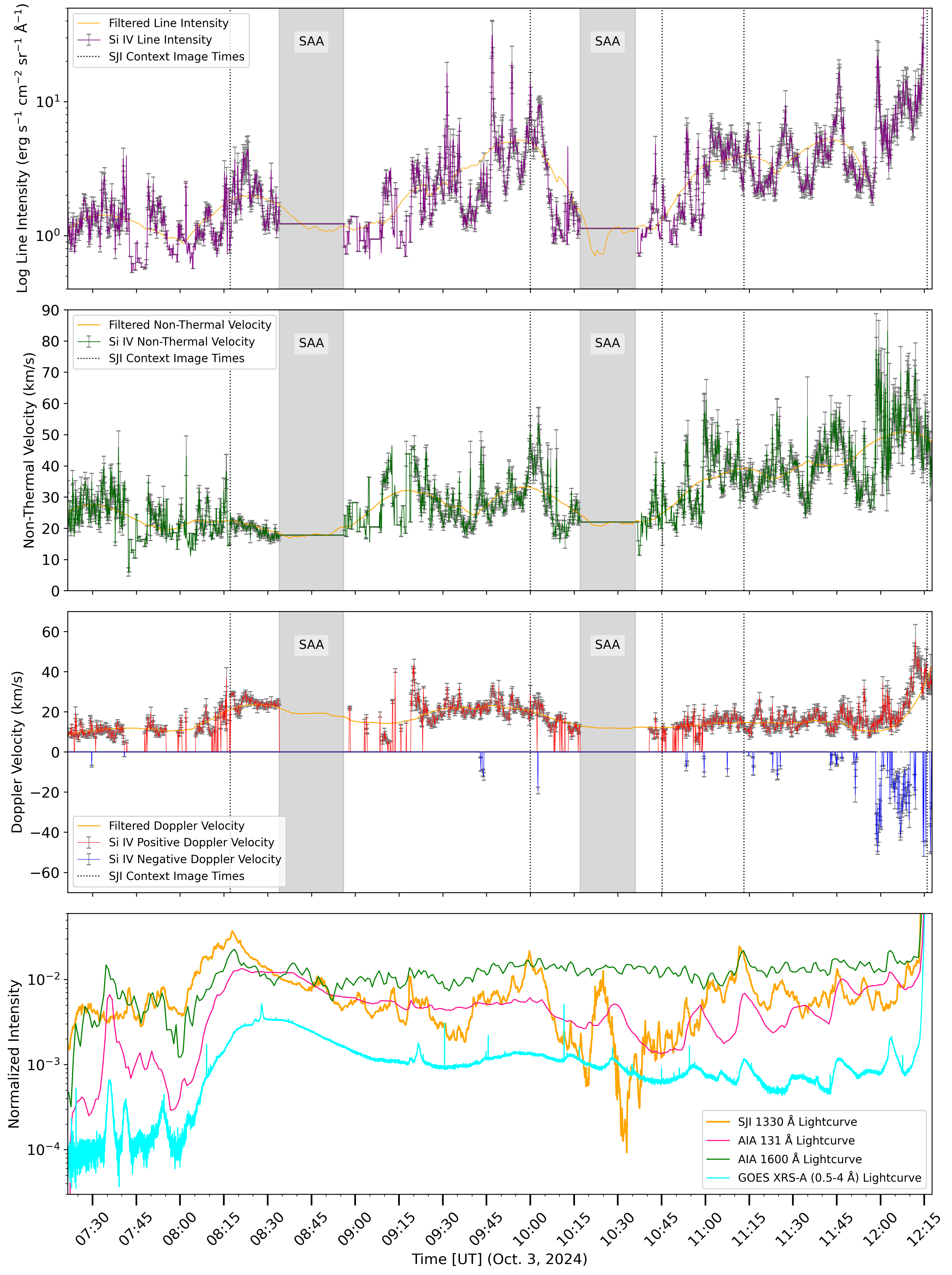}}
\small
\caption{\textit{Top three rows:} Time series of three quantities derived from the Si~IV line of IRIS averaged at each time-step over 24 rebinned pixels around the PIL: line intensity (\textit{first row}), non-thermal velocity (\textit{second row}), and Doppler velocity (\textit{third row}). The grayed-out areas (marked with 'SAA') correspond to the passing of the satellite through the South-Atlantic Anomaly, corrupting data from IRIS during these time intervals.
\textit{Bottom row:} Lightcurves from GOES 0.5--4 \AA\ (cyan), SJI 1330 \AA{} (orange), AIA 131 \AA\ (pink), and AIA 1600 \AA\ (green).
See \S~\ref{temp_res} for more details.}
\label{1D}
\end{figure}

The top row of Figure \ref{sji} shows five snapshots of the flare evolution (SJI 1330 \AA{}). Beneath this row, we present time-distance maps of the Si IV line intensity, non-thermal velocity, and Doppler velocity quantities in the upper part of the slit (as highlighted in Figure \ref{cmaps}) during the five hours preceding the flare.
This portion of the slit, spanning a 24-pixel region (after binning), crosses the central part of the active region's PIL.
The log-scaled intensity map (second row) reveals that values near the PIL far exceed the background values. These maps illustrate the spatial development of the observed features, particularly in the final hour before the flare. Doppler blueshifts emerge predominantly on both sides of the PIL near the footpoints, suggesting upflows aligned with magnetic field lines undergoing reconnection. These regions coincide spatially and temporally with localized peaks in non-thermal velocity.

Figure~\ref{1D} presents the temporal evolution of the averaged (across 24 rebinned pixels) Si IV line intensity, non-thermal velocity, and Doppler velocity quantities in the same slit region as Figure \ref{sji} during the five hours preceding the flare. Each quantity is plotted with gray error bars representing standard errors and a Savitzky–Golay smoothed trend (orange line).
The shaded ``SAA regions" at 08:33-08:55 UT and 10:17-10:35 UT correspond to data collected while IRIS was passing into the South Atlantic Anomaly (SAA), which were removed due to low data quality.
The bottom panel includes lightcurves from the GOES 0.5-4 \AA\ channel, SJI 1330 \AA, AIA 131 \AA\ and AIA 1600 \AA.

From the second panel of Figure~\ref{1D} we find that the non-thermal velocity exhibits a clear increasing trend, rising from $\sim$20–30 km\,s$^{-1}$ between 07:30 - 08:30 UT to $\sim$40–50 km\,s$^{-1}$ by 11:00 UT, and up to $\sim$60 km\,s$^{-1}$ at 12:00 UT. The non-thermal velocity also displays complex, possibly quasi-periodic variations that we further analyze in \S~\ref{peri_res}.

We find that the Doppler velocity (Figure~\ref{1D}, third panel) remains predominantly positive, indicating persistent downflows. To isolate physical plasma motions, we separate and analyze positive (redshift) and negative (blueshift) values separately, filtering out unreliable times with insufficient pixel representation. Prior to 11:00 UT, redshifts ($\sim$10–30 km\,s$^{-1}$) dominate.
After 11:00 UT, a notable rise in blueshifts is observed, reaching $20$–$40$ km\,s$^{-1}$ by 12:00 UT and exceeding the redshifts in amplitude, consistent with enhanced upflows or chromospheric evaporation 15 minutes before the flare start time. This transition also coincides with the rise in non-thermal velocity.

The line intensity in Figure~\ref{1D} (first panel) follows an upward trend similar to Doppler velocity, with elevated values between 09:00 - 10:00 UT, a dip around 10:05 UT, and a renewed increase in the final hour before flare onset. Its fluctuations are more strongly modulated, showing periodic signatures.

\subsection{Wavelet Analysis (5 Hours Pre-Flare)}\label{wave_res}

\begin{figure}
\centerline{\hspace*{0.015\textwidth}
    \includegraphics[width=0.97\textwidth]{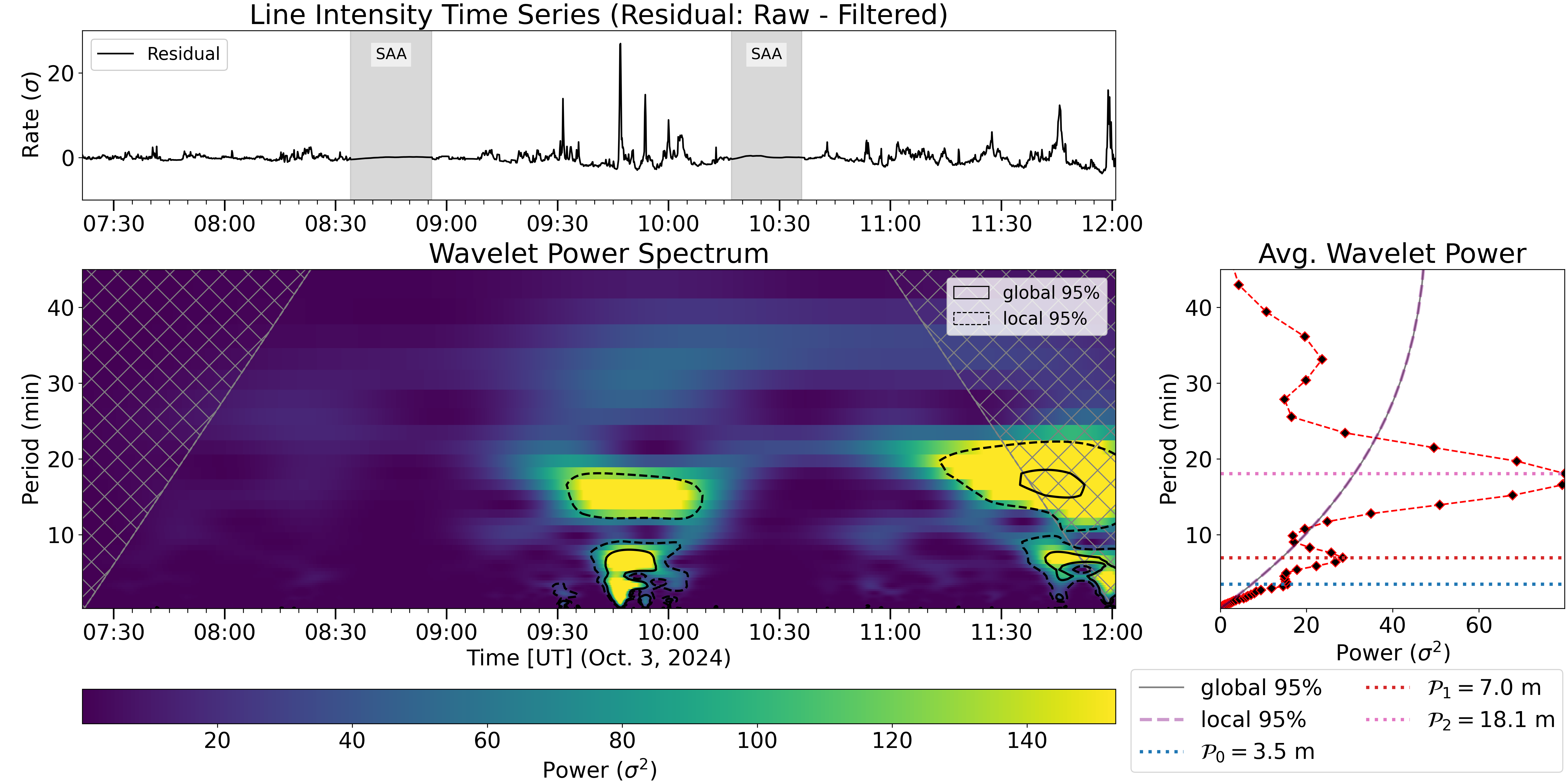}
}
\vspace{-0.32\textwidth}
\centerline{\Large \bf \hspace{0.0\textwidth}}
\vspace{0.30\textwidth}

\centerline{\hspace*{0.015\textwidth}
    \includegraphics[width=0.97\textwidth]{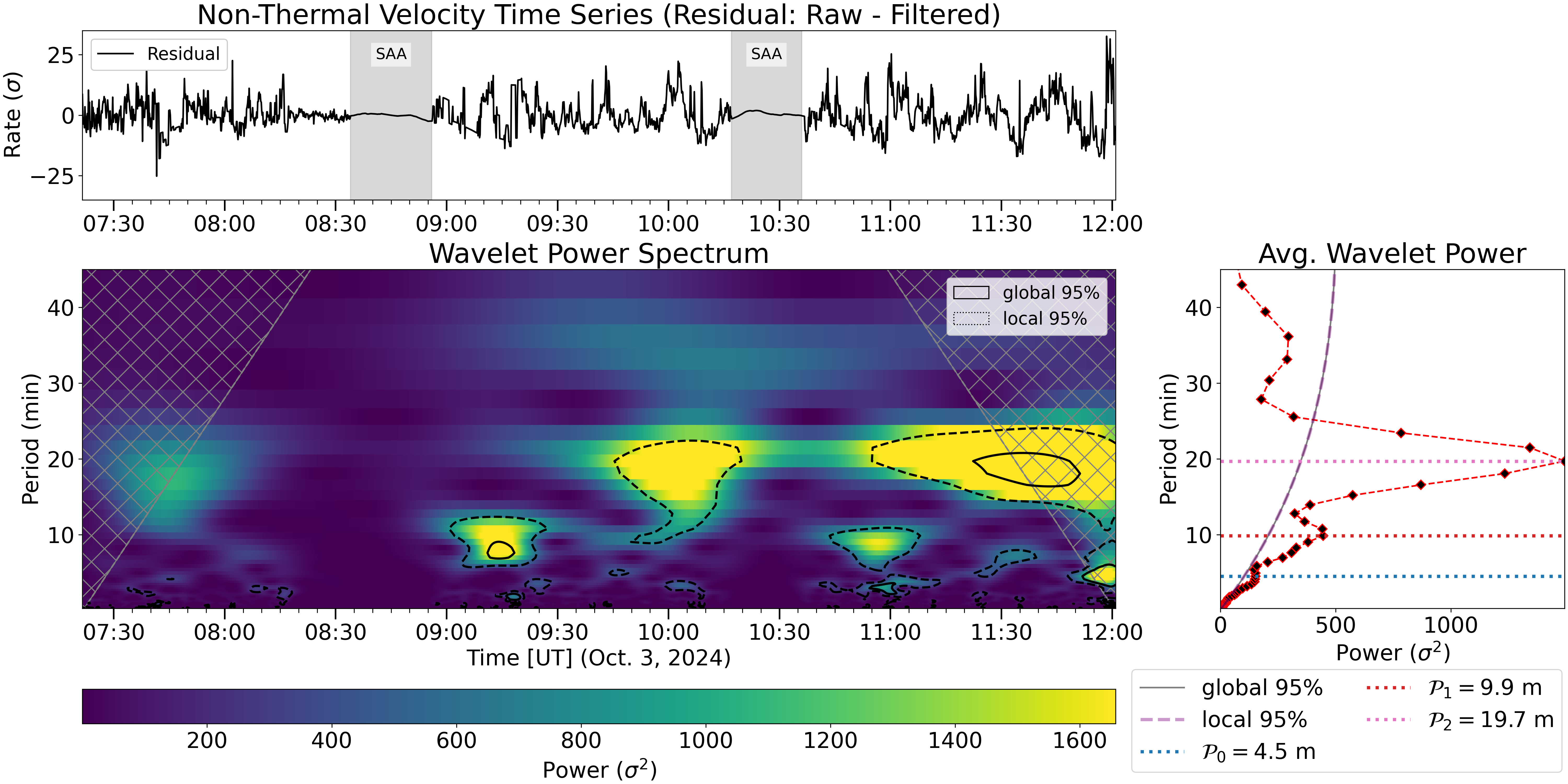}
}
\vspace{-0.32\textwidth}
\centerline{\Large \bf \hspace{0.0\textwidth}}
\vspace{0.30\textwidth}

\centerline{\hspace*{0.015\textwidth}
    \includegraphics[width=0.97\textwidth]{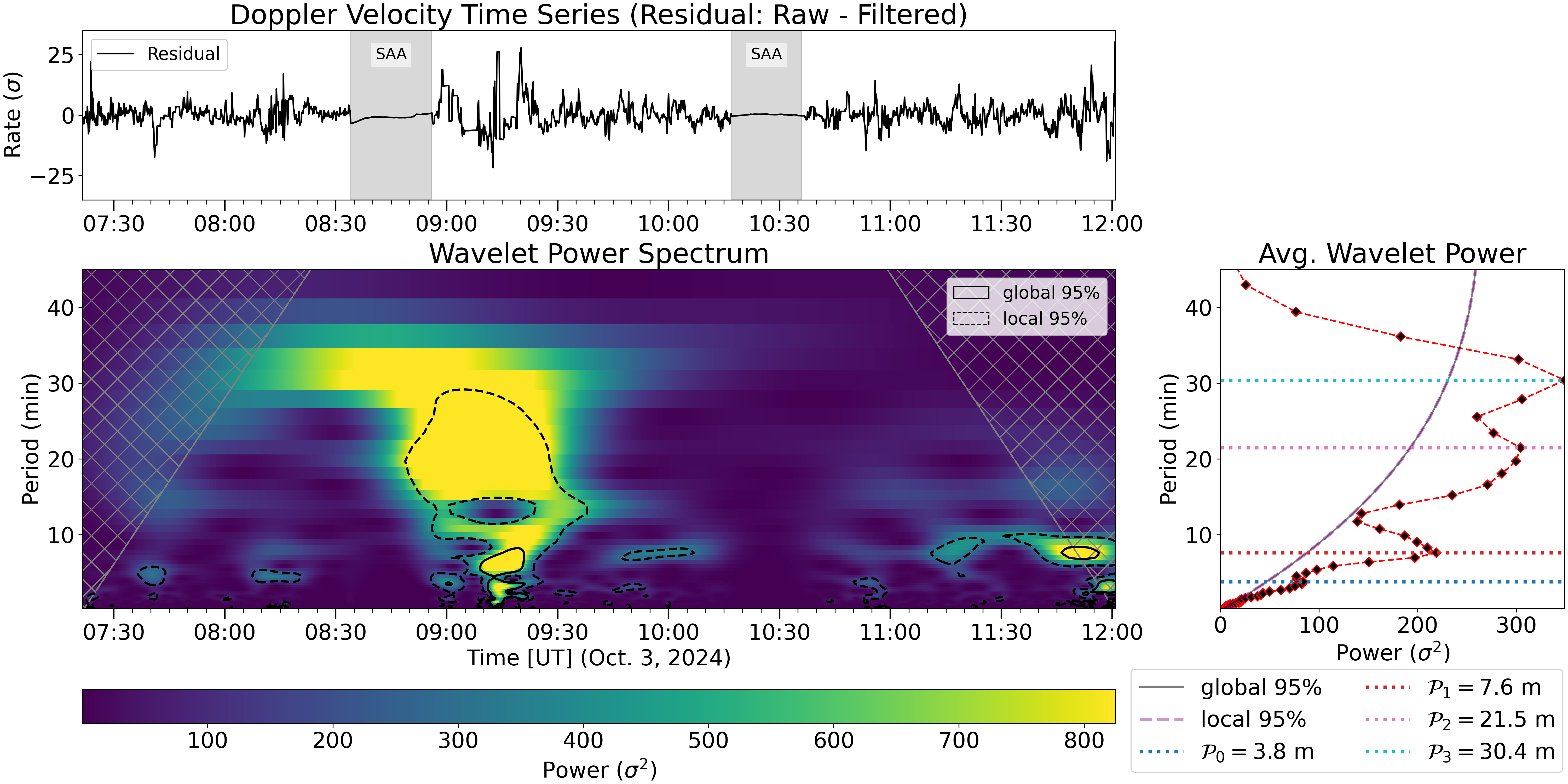}
}
\vspace{-0.32\textwidth}
\centerline{\Large \bf \hspace{0.0\textwidth}}
\vspace{0.30\textwidth}

\small
\caption{Residual time series and wavelet power spectra for line intensity (top), 
non-thermal velocity (middle) and Doppler velocity (bottom) derived from the IRIS Si~IV line. 
For each parameter:
(Top Left) Residual time series after subtracting a smoothed trend given by the Savitzky–Golay filter. 
(Bottom Left) Wavelet power spectrum. The cross-hatched areas represent the cone of influence, 
where edge effects become significant. Solid and dashed red contours mark the global and local 95\% 
confidence levels, respectively.
(Bottom Right) Global wavelet spectrum averaged over time, with horizontal dotted lines indicating 
multiple detected peak periods. The brown and gold curves show the global and local 95\% 
significance thresholds.
See \S~\ref{wave_res} for more details.}
\label{wavelets}
\end{figure}

In this section, we analyze periodic behavior in the Si IV non-thermal velocity, Doppler velocity, and line intensity over five hours before the flare. We use the wavelet analysis as described in \S~\ref{methods}.
In Figure \ref{wavelets} we present three panels per quantity: the detrended time series (top left), constructed by subtracting the Savitzky–Golay smoothed trend from the raw data (see \S~\ref{methods}); the wavelet power spectrum (bottom left), with 95\% global and local significance contours and the cone of influence indicated by the cross-hatched region; and the time-averaged wavelet power spectrum (right). The analysis spans from 07:21 to 12:00 UT, ending 15 minutes before the flare start time to avoid contamination from the rapid intensity rise. Peak periods representing the strongest $30$\% of the power distribution are highlighted with dashed lines in the averaged spectrum to isolate dominant oscillatory behavior.

In Figure \ref{wavelets} (middle), the non-thermal velocity series exhibit multiple oscillations throughout the observing window, most notably during the final hour before the flare (11:00–12:00 UT). The averaged wavelet power reveals enhanced energy at periods near 20, 10, and 5 minutes, all exceeding the confidence thresholds but more significantly for the 20 and 10 minute peaks. The signal around 20 minutes emerges between 09:45 and 10:15 UT, gradually weakens until about 11:15 UT, and subsequently strengthens beyond its initial amplitude. Shorter-period fluctuations near 10 minutes are more temporally localized, appearing intermittently around 09:15 and 11:00 UT.

In Figure \ref{wavelets} (top), line intensity variations display a comparable pattern to the non-thermal velocity series, with significant power concentrated near 20 and 8 minutes and weaker contributions at longer timescales. 
The temporal evolution of these features also broadly coincides, suggesting that both quantities respond to the same oscillatory driver.
The line intensity power spectrum reveals a dominant signal near 18 minutes, active during 09:30–10:15 UT and again from 11:30 to 12:00 UT, accompanied by weaker short-period components around 7 minutes.

In Figure \ref{wavelets} (bottom), the Doppler velocity fluctuations are more confined in time and show enhanced amplitudes, particularly around 09:15 UT. 
Additionally, a 30 minutes peak appears through this quantity, focused between 8:30 and 09:30 UT with the highest power signature for this wavelet.

Taken together, these wavelets indicate two characteristic ranges of oscillatory behavior: a long-period component around 18–22 minutes, and a shorter-period component near 8–10 minutes, which appear across all observables.
An even shorter period regime ($\approx$4 minutes) is also present in all wavelets but likely corresponds to the well-documented 3-minute oscillation in the chromosphere and transition region \cite[e.g.][]{Bogdan2006}.

We find comparable periodicities present in AIA and GOES data, including $\sim$17 min in AIA 131, $\sim$5 and 22 min in AIA 1600, and $\sim$6 min in GOES XRS-A. However, these signatures primarily occur during the $\sim$90 minutes preceding the flare, whereas IRIS pre-flare signatures extend to over $\sim$3 hours before the flare. We suggest that the longer-duration pre-flare signatures are detected only in IRIS due to its higher sensitivity and resolution. The IRIS observations therefore reveal sustained 3-hour pre-flare activity beyond the shorter 90-min interval seen in the other datasets.

\subsection{Evolution during the late pre-flare phase (1-Hour Pre-Flare)}\label{peri_res} 

\begin{figure}
\centerline{\includegraphics[width=\textwidth]{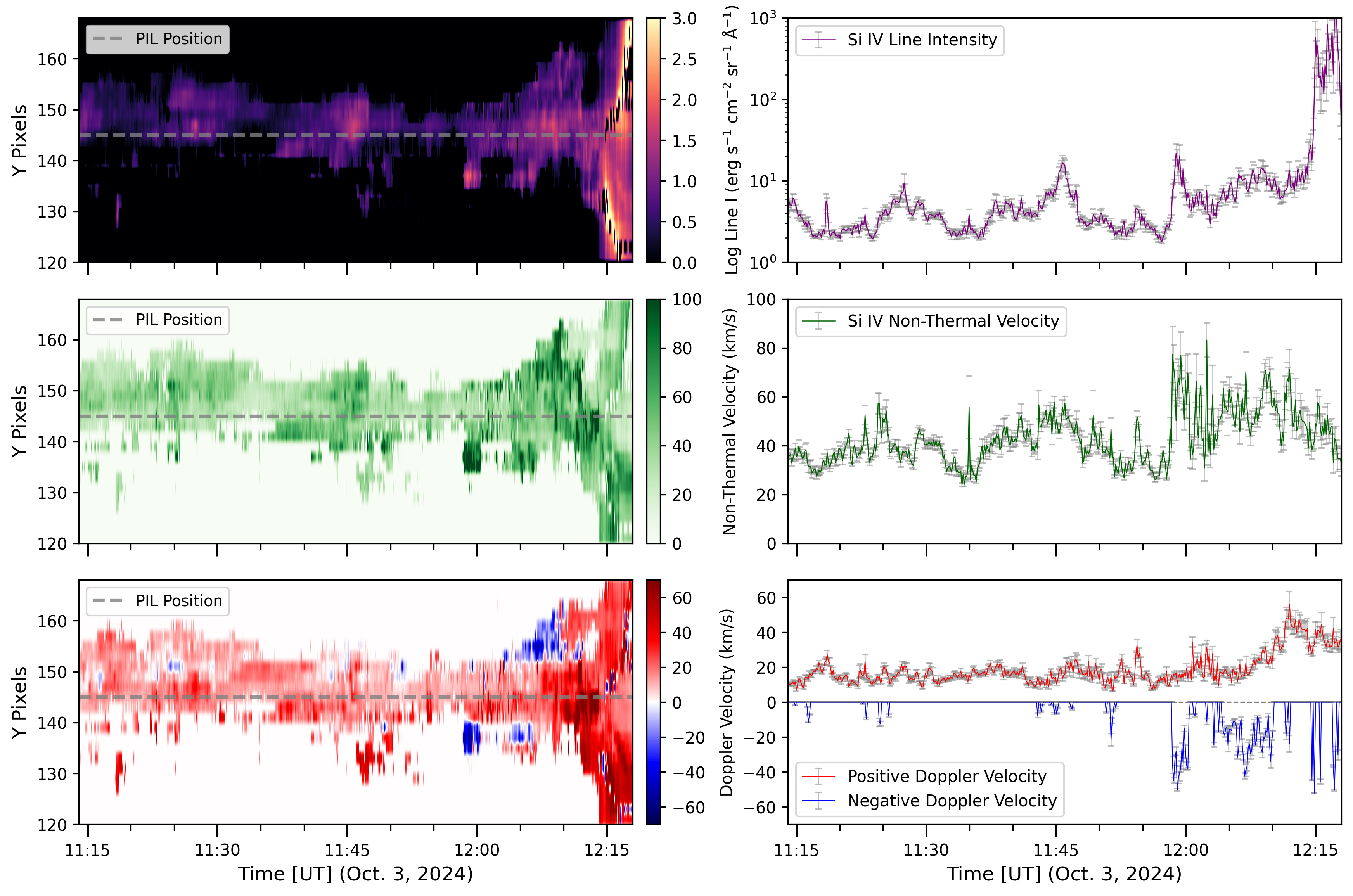}}
\small
\caption{Subplot of Figure \ref{1D}, showing one-hour of pre-flare evolution. Panels from top to bottom: line intensity, non-thermal velocity, Doppler velocity, derived from the Si~IV line of IRIS.
\textit{Left side:} Time-distance maps spanning over 48 pixels of the IRIS slit on the PIL. The line intensity map is logged.
\textit{Right side:} Time series averaged for the same pixels at each time-step.
See \S~\ref{peri_res} for more details.}
\label{zoom}
\end{figure}

Figure \ref{zoom} presents a cropped version of the time-distance maps of Figure \ref{sji} and of the time series of Figure \ref{1D}, focusing on the last hour of Doppler velocity, non-thermal velocity and intensity evolution from 11:15 - 12:15 UT, before flare onset at 12:15 UT. 
While wavelet analysis of 5 hours of pre-flare observations (see Figure \ref{wavelets}) confirmed the presence of periodicities in these three Si IV parameters, their temporal relationship to one another becomes clearer in the hour preceding the flare. We observe peaks in non-thermal velocity and line intensity coinciding at approximately 11:27 UT, 11:45 UT, and 12:00 UT, consistent with the $\sim19.2$-minute and $18.1$-minute period signatures identified in 5-hr dataset analysis in Figure \ref{wavelets}. These oscillations are especially evident in the line intensity curve and are less pronounced but still discernible in the non-thermal velocity, which also follows a broader upward trend.

A significant transition occurs at $\sim$11:58 UT, 17 minutes before the flare, marked by an abrupt increase in all three quantities. The non-thermal velocity jumps from $\sim30$ to over $70$ km\,s$^{-1}$, the Si IV intensity rises by nearly an order of magnitude, and Doppler blueshifts intensify to $-50$ km\,s$^{-1}$, while redshifts remain largely unchanged. Notably, a C4.6 flare occurred in this active region from 11:57 UT to 12:15 UT (peaking at 12:04 UT).

In addition to the overall increase over 1-hr before the flare, we find that the intensity, non-thermal velocity, and Doppler velocity increase during the final 16 minutes before the flare start time at 12:15 UT. This increase is aligned spatially and temporally between the three parameters. 

To further quantify these relationships, we performed a cross correlation analysis of the normalized time series of Si IV intensity, non-thermal velocity and Doppler velocity, in the 60 minutes preceding the flare. We find a strong in-phase relationship between intensity and non-thermal velocity, with Spearman rank coefficient r$_s=$ 0.76. 
In contrast, we find Doppler velocity to be out of phase with intensity and non-thermal velocity, with a maximum correlation coefficient of r$_s=$ 0.43 and r$_s=$ 0.37 respectively, both at a lag time of $\sim8-9$ minutes. Considering our identification of 18-21 minutes in the wavelet analysis, a peak correlation at $\sim8-9$ minutes corresponds to an out-of-phase relationship between Doppler velocity with intensity and non-thermal velocity. We note that because the Doppler velocity remains redshifted for most pixels, we are observing a drop in downflow velocity with rising intensities and non-thermal velocities.

\section{Discussion}\label{diss}

We analyze the spectra of the Si IV 1403 \AA\ IRIS observations for five hours preceding flare onset, finding an increase in intensity, Doppler and non-thermal velocities in the chromosphere/transition region up to 3 hours before the flare, with the greatest increase in the final 15-20 minutes.
These changes in behavior occur at and around the PIL area of the active region, with redshifts dominating during the 3-hour period but blueshifts appearing on either side of the PIL during the last 20-min of the pre-flaring period. Integrating intensity, Doppler and non-thermal velocities separately within and outside the PIL area along the IRIS slit, we find that throughout most of the pre-flare period, these quantities are dominated by regions outside of the PIL area along the slit. However, approximately 30 minutes prior to the flare, the dynamics become primarily dominated by the PIL.
From wavelet analysis in the Si IV parameters, we also find persistent periodic oscillations during the $3$ hours preceding flare onset, with dominant periods occurring in the $\sim7-10$-minute and $\sim18-21$-minute ranges.

Previous pre-flare studies using Hinode/EIS and IRIS have reported increases in non-thermal line broadening and Doppler shifts minutes to an hour before flare onset \citep[e.g.][]{Woods_preflaring,Harra_2009,Harra2013,Wallace2010,Imada2014,Bamba_2017,Rubio2012,To2025}. Most IRIS Si IV analyses \citep[e.g.][]{Woods_preflaring, Bamba_2017} employed raster observations, providing broad spatial context but limited temporal resolution. In our work, however, we use sit-and-stare observations focusing on a single spatial slice across the active region PIL with rapid cadence over a $5$-hour period, allowing us to track the long-duration temporal evolution of pre-flaring plasma.
The gradual rise in non-thermal velocities we observe is broadly consistent with earlier reports of localized pre-flare enhancements at loop footpoints and looptops \citep{Harra2013} and with the Si IV blueshifts identified tens of minutes before flares by \citet{Woods_preflaring}. While differences in how Doppler shifts are measured prevent direct quantitative comparison, the qualitative agreement is clear.

More importantly, our results reveal much longer, up to three-hour, periods of progressive increase in non-thermal velocity prior to this X-class flare. In light of recent works \citep[e.g.][]{Harra2013,Woods_preflaring,To2025}, we interpret these extended precursors as evidence of a slow, large-scale destabilization of the coronal magnetic configuration. Such destabilization may reflect the gradual activation of a flux rope, which can potentially proceed through a magnetic avalanche (e.g., \citealt{Chitta2025}). In magnetic avalanche, small reconnection events cascade through the system. 3D MHD simulations \citep{Cozzo2023,Cozzo2024} show that these processes naturally produce enhanced non-thermal broadening. Our observed long-duration increases may therefore represent the spectroscopic signature of this progressively more energetic reconnection events leading up to eruption.
 
Superimposed on this long-term evolution is a distinct transition in the final $15$-$20$ minutes before the flare. During this period, we detect abrupt increases in non-thermal velocities (see Figure \ref{zoom}), enhanced intensities, and the appearance of localized blueshifts. These signatures point to a shift from slow evolution to more fast energy conversion. The enhanced non-thermal velocity indicates increased turbulence or wave-driven broadening \citep{Harra2013}, while blueshifted Si IV emission reflects upward motions from the transition region toward the corona, resembling chromospheric-evaporation-like flows \citep[e.g.][]{Bamba_2017,milligan2009,young2013}. This sharp intensification in pre-flare dynamics leads to flare onset.

The statistical study of \citet{To2025} provides context for these results, showing that X-class flares exhibit the longest precursor detection times (up to 48 minutes) in coronal Fe XVI lines observed with Hinode/EIS. Their measurements sample 1-2 MK plasma within coronal loops, whereas the Si IV line used here forms in the lower transition region at $\approx8\times10^4$K and primarily traces flare ribbons and footpoints. A direct quantitative comparison between EIS- and IRIS-derived non-thermal velocities is therefore not meaningful, but the qualitative agreement in the timing of enhanced dynamics across such distinct atmospheric layers suggests that pre-flare energy release may span multiple heights within the active region.

Several physical processes may contribute to the observed pre-flare evolution of Doppler velocity.
The dominance of redshifts in the early phase (see Figure \ref{sji}) could be associated with low-level reconnection or chromospheric condensation driven by gradual energy deposition \citep{Fisher1985}. 
As the pre-flare energy accumulation progresses, the strong blueshifts may indicate the transition to upward evaporative flows, driven by intensified magnetic reconnection and energy deposition in the lower atmosphere \citep{Fisher1985, Tian2018}.
\citet{Sadykov_2015} reported redshifts in the C II k line of an M-class flare before its impulsive phase, which they interpreted as plasma falling back along magnetic loops rather than as part of the chromospheric evaporation process.

Finally, we detect persistent periodic signatures across Si IV quantities in intensity, Doppler velocity, and non-thermal velocity throughout the three-hour pre-flare interval. The alignment of dominant periods across all three parameters points to a common driver, such as magnetoacoustic or Alfvénic oscillations. 
The quasi-periods detected may represent harmonics of a common oscillatory process, further supporting a wave-based interpretation of the pre-flare activity.
Very-long-period pulsations of similar duration were statistically observed in X-class pre-flare intervals by \citet{Tan2016}. Their study found pulsating periods averaging at $16.1\pm10.7$ minutes for an X-class flare population of 18 flares, that lasted for $85.6\pm33.7$ minutes before flare onset, which is comparable to our $\sim18-21$ minutes signature. 
Periods corresponding closely to our faster ranges of pulsations ($\sim7-10$ minutes) were found preceding an X-class flare by \citet{Zimovets_2025}, who report $\sim5-8$ minutes periods of pre-flare QPPs from X-ray data. They interpret these pulsations as a series of magnetic reconnection events, consistent with our magnetic avalanche scenario.
Meanwhile, recent data-constrained MHD simulations of the same active region \citep{Matsumoto2025} show multi-hour magnetic energy buildup and restructuring. Such a slowly evolving environment provides a natural setting for wave activity to modulate pre-flare plasma properties, as we observe.

\section{Conclusion}\label{conc}

We analyze IRIS Si IV 1403 \AA\ spectra of AR 13842 for 5 hours before an X9.0-class flare on October 3, 2024, to investigate dynamics of the pre-flare phase.
We present 1D and 2D time-series evolution and conduct wavelet analysis of intensity, Doppler velocity, and non-thermal velocity measurements along the sit-and-stare slit centered around the PIL region. Our key findings are:
\begin{enumerate}
    \item From 3 hours preceding flare onset, Si IV line intensity, Doppler shift, and non-thermal velocity all show gradual increases (Fig. \ref{1D}), pointing to a steady accumulation of energy in the transition region.
    \item At 11:58 UT ($\approx$ 20 minutes before the flare), we measure a notable rise in non-thermal velocity, pronounced blueshifts, and enhanced intensity (Fig. \ref{zoom}), marking a clear shift in pre-flare dynamics. These changes possibly arise from localized reconnection or chromospheric evaporation.
    \item Examining the location of Si IV properties along the slit reveals cospatial blueshifts and non-thermal enhancements centered near the PIL (Fig. \ref{zoom}), showing the importance of reconnection-driven plasma dynamics early in the flare's life.
    \item Finally, wavelet analysis reveals consistent periodicities during the 3 hours preceding flare onset across the Si IV parameters, with dominant peaks at both $\sim7-10$ min. and $\sim18-21$ min. ranges (Fig. \ref{wavelets}). Within the final hour before the flare, we see a clear in-phase relationship between intensity and non-thermal velocity, both in anti-phase with Doppler velocity. The alignment and correlation between non-thermal velocity and line intensity oscillations suggests the presence of wave activity during the pre-flare phase.
\end{enumerate}

This work presents unique IRIS observations of five hours of high-cadence pre-flare observations of a major solar flare. 
We reveal insights into the timescales to which the pre-flare phase begins, providing indications for early reconnection and wave dynamics during this phase of the flare -- consistent with the slow destabilization of the solar atmosphere prior to flare onset. 
A better understanding of the pre-flare phase, backed up by theory and simulation work, is needed before we can fully understand the energy release processes in solar flares.

\section*{Acknowledgments}
L.S. acknowledges support from the Observatory of Paris - PSL Graduate Program (IRT). 
M.K.D. acknowledges support from NASA ECIP NNH18ZDA001N and NSF CAREER SPVKK1RC2MZ3 awards. 
R.J.F. thanks support from NA\-SA HGI award 80NSSC25K7927.

\bibliographystyle{aasjournal}
\bibliography{biblio}

\end{document}